\documentclass[unsortedaddress,prd]{revtex4-1}

\usepackage{amsmath}
\usepackage{amsfonts}
\usepackage{amsthm}
\usepackage{amssymb}
\usepackage{graphicx}
\usepackage{hyperref}

	\newcommand{\ncd}{\newcommand}
	\ncd{\mrm}    {\mathrm}
	\ncd{\beq} {\begin{equation}}
	\ncd{\eeq} {\end{equation}}

	\def\d{{\rm d}}

	\newcommand{\dlie}{\mathcal{L}}

	\newcommand{\cp}[1]{p^{#1}}
	\newcommand{\cx}[1]{q^{#1}}

	\newcommand{\lie}{\mathcal{L}}

	\newcommand{\prtf}[2]{\frac{\partial#1}{\partial #2}}

\begin{document}

	\title{Infinitesimal Legendre symmetry in the Geometrothermodynamics programme}
	
		\author{D. Garc\'ia-Pel\'aez}
		\email{dgarciap@up.edu.mx}
		\affiliation{Instituto de Ciencias Nucleares, Universidad Nacional Aut\'onoma de M\'exico,\\
A.P. 70-543, 04510 M\'exico D.F., M\'exico}
		\affiliation{Universidad Panamericana,\\ Tecoyotitla 366.
Col. Ex Hacienda Guadalupe Chimalistac, 01050 M\'exico D.F., M\'exico}

		\author{C. S. L\'opez-Monsalvo}
		\email{cesar.slm@correo.nucleares.unam.mx}
		\affiliation{Instituto de Ciencias Nucleares, Universidad Nacional Aut\'onoma de M\'exico,\\
A.P. 70-543, 04510 M\'exico D.F., M\'exico}

	\begin{abstract}
The work within the Geometrothermodynamics programme rests upon the metric structure for the thermodynamic phase-space. Such structure exhibits discrete Legendre symmetry. In this work, we study the class of metrics which are invariant along the infinitesimal generators of Legendre transformations.  We solve the Legendre-Killing equation for a $K$-contact general metric. We consider the case with two thermodynamic degrees of freedom, i.e. when the dimension of the thermodynamic phase-space is five. For the generic form of contact metrics, the solution of the Legendre-Killing system is unique, with the sole restriction that the only independent metric function -- $\Omega$ -- should be dragged along the orbits of the Legendre generator. We revisit the ideal gas in the light of this class of metrics. Imposing the vanishing of the scalar curvature for this system results in a further differential equation for the metric function $\Omega$ which is not compatible with the Legendre invariance constraint. This result does not allow us to use the regular interpretation of the curvature scalar as a measure of thermodynamic interaction for this particular class.

	\end{abstract}

\maketitle

\section{Introduction}

The mathematical nature of thermodynamics has been largely explored over the past few decades. Motivated by the early works on metric formulations for thermodynamics of Rao and Fisher \cite{fisher}  together with  the later developments in contact geometry applied to thermodynamics of Mrugala \cite{mrugala1, mrugala2}, the central goal of the  GTD programme  has been to provide a Legendre covariant description of thermal phenomena \cite{quevedo}. However, a consistent way for obtaining Legendre invariant metrics whose associated geometry reveals some physical property has been lacking since the birth of the programme. The difficulty of such a task lies in the fact that \emph{discrete} Legendre transformations do not form a group. A summary of the various challenges for the formalism can be found in \cite{conformal,conformalII}.

The metric based programme for geometric thermodynamics presented here considers Legendre symmetry as a fundamental building block for the formalism. From a physical point of view, this is consistent with the fact that in an experiment certain thermodynamic quantities may not be suitable for direct measurement, e.g. in a system with two degrees of freedom represented by the molar internal energy, where the molar entropy and volume are the relevant variables.  In such cases, one can use `conjugate'  variables, e.g. temperature and pressure, which are easier to control without changing the physical content of the conclusions one can reach from such experiment. This is an indication that thermodynamics is independent of the potential used to describe a particular phenomena. Therefore, in any metric theory of thermodynamics, if one assigns some physical significance to quantities derived from the metric tensor, one should demand Legendre covariance. In this respect, the GTD programme has been successful. 

 A natural way to set up a geometric theory of thermodynamics has been  to endow the working space with a metric. It is at this point that different geometric programmes for thermodynamics diverge. On the one hand, there is the fluctuation based formalism of Ruppeiner \cite{rupp}, where the metric tensor  and its associated line element account for the probability of fluctuations. On the other hand, the GTD programme equips the thermodynamic phase-space (c.f. Section \ref{secII}) with a Legendre invariant metric which is carried over to the equilibrium space by means of an embedding map. This guarantees that the curvature of the equilibrium space is also preserved by a Legendre transformation, that is, independent of the thermodynamic potential. However, its relationship with thermodynamic fluctuation theory is a bit less clear. Finally, let us note that the metrics used within the GTD programme, albeit being manifestly invariant under discrete Legendre transformations, they are not invariant along the orbits of the generating vector field of such transformations.  This should be contrasted with the case of Special Relativity, where the flat Minkowski metric is Lorentz invariant and is carried along the orbits of the infinitesimal generators of the Lorentz group. \cite{stewart}

In this work, we explore the infinitesimal generators of the Legendre transformations and construct a family of metrics whose isometry group contains these generators. This turns out to be a very strong constraint on the possibilities for Legendre invariant metrics. In particular, none of the metrics used within the GTD programme satisfies such a constraint. 

We solve the system defined by the Killing equation associated with the generators of Legendre transformations for an arbitrary metric in the thermodynamic phase-space with the sole restriction of being a K-contact, i.e. that the Reeb vector field associated with the choice of contact form generates an extra symmetry. We perform the  analysis by considering a system with two thermodynamic degrees of freedom which can be easily extended to the general case.  

The paper is organised as follows. In section  \ref{secII}, we provide a brief introduction to the geometry of thermodynamics and the general framework of the  GTD programme. In section III, we introduce the contact Hamiltonians as the generating functions for the infinitesimal generators of Legendre transformations. This approach to Legendre symmetry in thermodynamics has been largely studied by Rajeev \cite{rajeev}. In section IV, we present a class of metrics whose isometries are given by infinitesimal Legendre transformations. We note that these metrics are \emph{conformally} associated with the contact structure of the thermodynamic phase space. Additionally, we show that none of the metrics used in the GTD programme belongs to this class. We explore the geometry of the space of equilibrium in the light of the induced metrics for this class and observe that we can no longer use the traditional interpretation for the curvature scalar. Finally, in section V we provide some closing remarks.

\section{The geometric structure of thermodynamics}
\label{secII}

Let us consider the standard  set-up (c.f. reference \cite{conformal}). To geometrise a thermodynamic system with $n$ degrees of freedom, we use a $2n+1$ dimensional contact manifold -- the thermodynamic phase-space $\mathcal{T}$  -- whose coordinates represent the thermodynamic variables. This allows us to take into account the First-Law of thermodynamics in a natural way through the maximal Legendre sub-manifold -- the equilibrium space $\mathcal{E}$ -- which is uniquely determined once the thermodynamic fundamental relation is known, i.e. when the thermodynamic potential has been fully specified.

Let us choose a contact 1-form  field -- $\eta \in T^*\mathcal{T}$ --  whose elements belong to the class generating the contact structure of the phase-space at each point. Such a field satisfies the non-integrability condition 
	\beq
	\eta \wedge \left(\d \eta \right)^n \neq 0.
	\eeq
Contact manifolds have the property that, in every patch  $\mathcal{U}\subset\mathcal{T}$  there is a set of local coordinates --Darboux's coordinates $\{\Phi,q^a,p_a \}$, with $a = 1...n$ -- such that at every point $x \in \mathcal{U}$ the contact 1-form can be written as
	\beq
	\eta = \d \Phi - p_a \d q^a,
	\eeq
where the Einstein's sum convention have been used. We will follow  such convention unless otherwise stated. 

The space of equilibrium states is the maximal dimensional integral sub-manifold embedded in $\mathcal{T}$ satisfying the isotropic condition
	\beq
	\varphi^*(\eta) = 0.
	\eeq
Here, the map $\varphi:\mathcal{E} \rightarrow \mathcal{T}$ is called a Legendre embedding. In Darboux coordinates, this is equivalent to the First-Law since
	\beq
	\varphi^*(\eta) = \varphi(\d \Phi - p_a \d q^a) = \left(\frac{\partial \Phi}{\partial q^a} - p_a \right) \d q^a = 0. 
	\eeq
The last equality serves as the definition of  the conjugate variables to the $q^a$'s through
	\beq
	\label{equil}
	\Phi= \Phi(q^a) \quad \text{and} \quad p_a = \frac{\partial \Phi}{\partial q^a}.
	\eeq
This is simply the well known statement that the fundamental relation $\Phi(q^a)$ completely specifies the thermodynamic system and, thus, the emebedding $\varphi$.

Additional to the contact structure, the GTD programme endows the thermodynamic phase-space with two families of metrics (c.f. section III in \cite{conformal}),
	\beq
	\label{quevedoGII}
	G_{\rm T}  = \eta \otimes \eta + \Omega \left (\xi^a_{\ b}q^b p_a \right) \left(\chi^c_{\ d}\ \d q^d \otimes \d p_c \right)
	\eeq
and
	\beq
	\label{quevedoGIII}
	G_{\rm P}  =\eta \otimes \eta +\Omega \sum_{i=1}^n \left[\left(q^i p_i \right)^{2k+1} \d q^i \otimes \d p_i \right].
	\eeq
Both have the discrete Legendre transformations as isometries, the former is only invariant under \emph{total} Legendre transformations and the latter is also invariant under partial Legendre transformations. We will clarify the role of Legendre symmetry in the GTD programme in the next section. Here, $\Omega$ is a Legendre invariant function of the Darboux coordinates,  $\xi^a_{\ b}$ and $\chi^a_{\ b}$ are $n \times n$ diagonal arrays of numbers providing the algebraic structure of each family and $k$ is an integer.

These families have been tested in various situations where the thermodynamic behaviour of a particular system is fully specified, allowing us to interpret the manifestly Legendre invariant curvature scalar associated with the induced metric in $\mathcal{E}$ as a ``measure of thermodynamic interaction'' (c.f. Quevedo \cite{quevedo}).

\section{Legendre Transformations and Contact Hamiltonians}

Legendre transformations are particular examples of contact diffeomorphisms, i.e.  symmetries of the contact structure of the thermodynamic phase-space. Notice that a transformation of $\mathcal{T}$ which leaves invariant its contact structure necessarily preserves the equilibrium space, since the tangent bundle of a Legendre sub-manifold is completely contained in the contact structure of $\mathcal{T}$ (c.f. Arnold \cite{arnold}). In this sense, a contact diffeomorphism of the thermodynamic phase-space, is a symmetry compatible with the First-Law. 

In thermodynamics, a Legendre transformation exchanges  the role of the extensive and intensive variables of a given fundamental function $\Phi(q^a)$. Consider the map $\phi:\mathcal{T} \rightarrow \mathcal{T}$ which leaves the contact structure invariant, that is, such that
	\beq
	\phi^*\eta = f\eta \quad \text{where} \quad f:\mathcal{T}\rightarrow \mathbb{R}^+.
	\eeq
The requirement that the function $f$ to be strictly positive simply means that we are considering only diffeomorphisms which preserve the orientation of the contact structure.

Let us call a differentiable function $h:\mathcal{T} \rightarrow \mathbb{R}$ a contact Hamiltonian. The Hamiltonian vector field generated by $h$ is defined through the relation
	\beq
	h = \eta\left[X_h\right].
	\eeq
In local Darboux coordinates it takes the form \cite{arnold}	
	\beq
	\label{generic.ham}
	X_h = \left( h - p_a \frac{\partial h}{\partial p_a}\right)\frac{\partial}{\partial \Phi} + \left(\frac{\partial h}{\partial q^a} + p_a \frac{\partial h}{\partial \Phi}\right)\frac{\partial }{\partial p_a} - \left(\frac{\partial h }{\partial p_a} \right)\frac{\partial }{\partial q^a}.
	\eeq	
Note that for every contact 1-form in the class defining the contact structure, there is a unique vector field $R_\eta$ such that
	\beq
	\eta[R_\eta] = 1 \quad \text{and} \quad \d\eta[R_\eta] = 0.
	\eeq
The vector field $R_\eta$ is called the Reeb vector field associated with the contact form $\eta$, and it is generated by the Hamiltonian $h=1$. In local Darboux coordinates, the Reeb vector field is simply
	\beq
	\label{reeb}
	R_\eta = \frac{\partial}{\partial \Phi}.
	\eeq

\begin{figure}
\includegraphics[width=0.35\columnwidth]{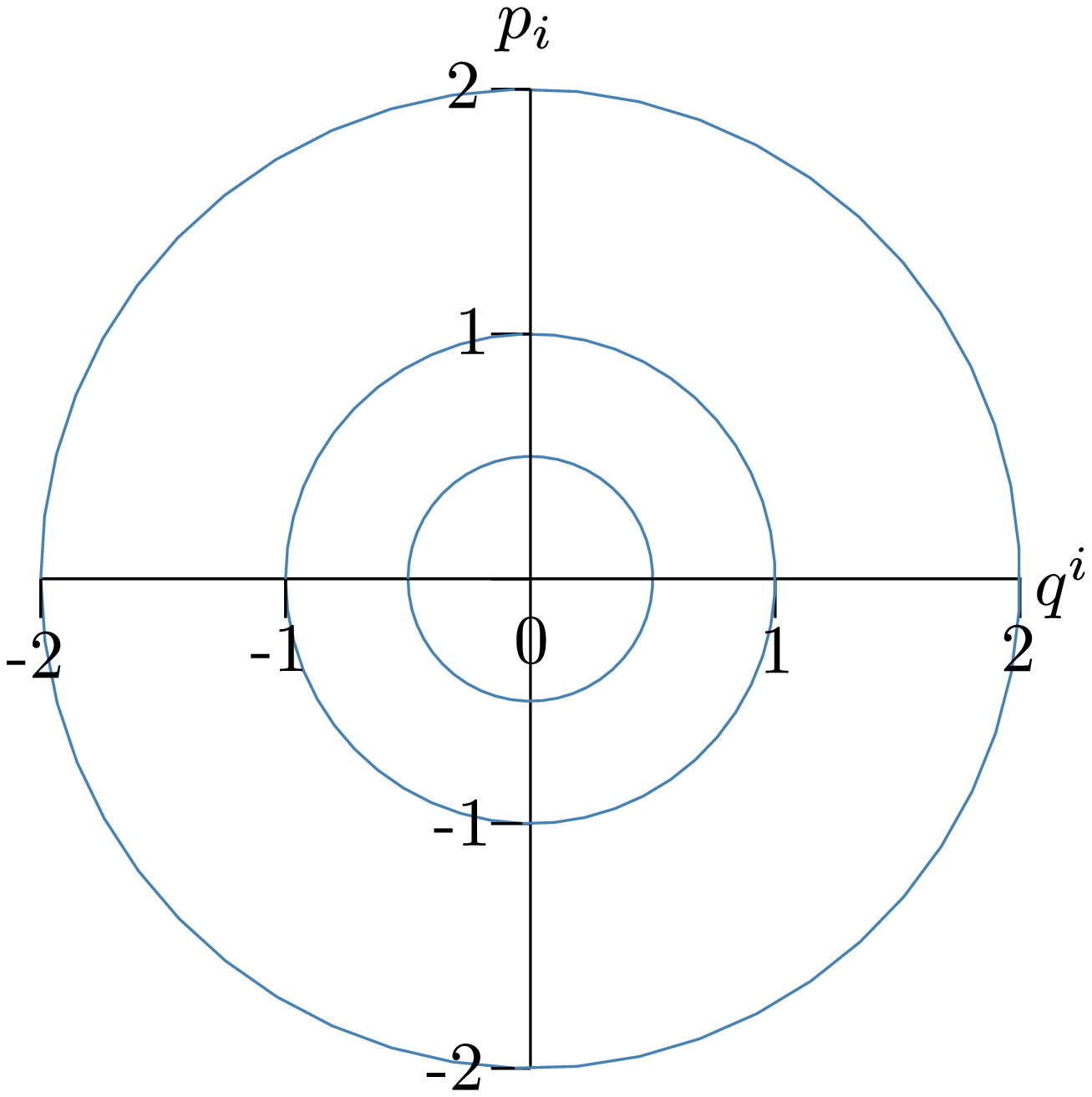}\hskip1cm\includegraphics[width=0.45\columnwidth]{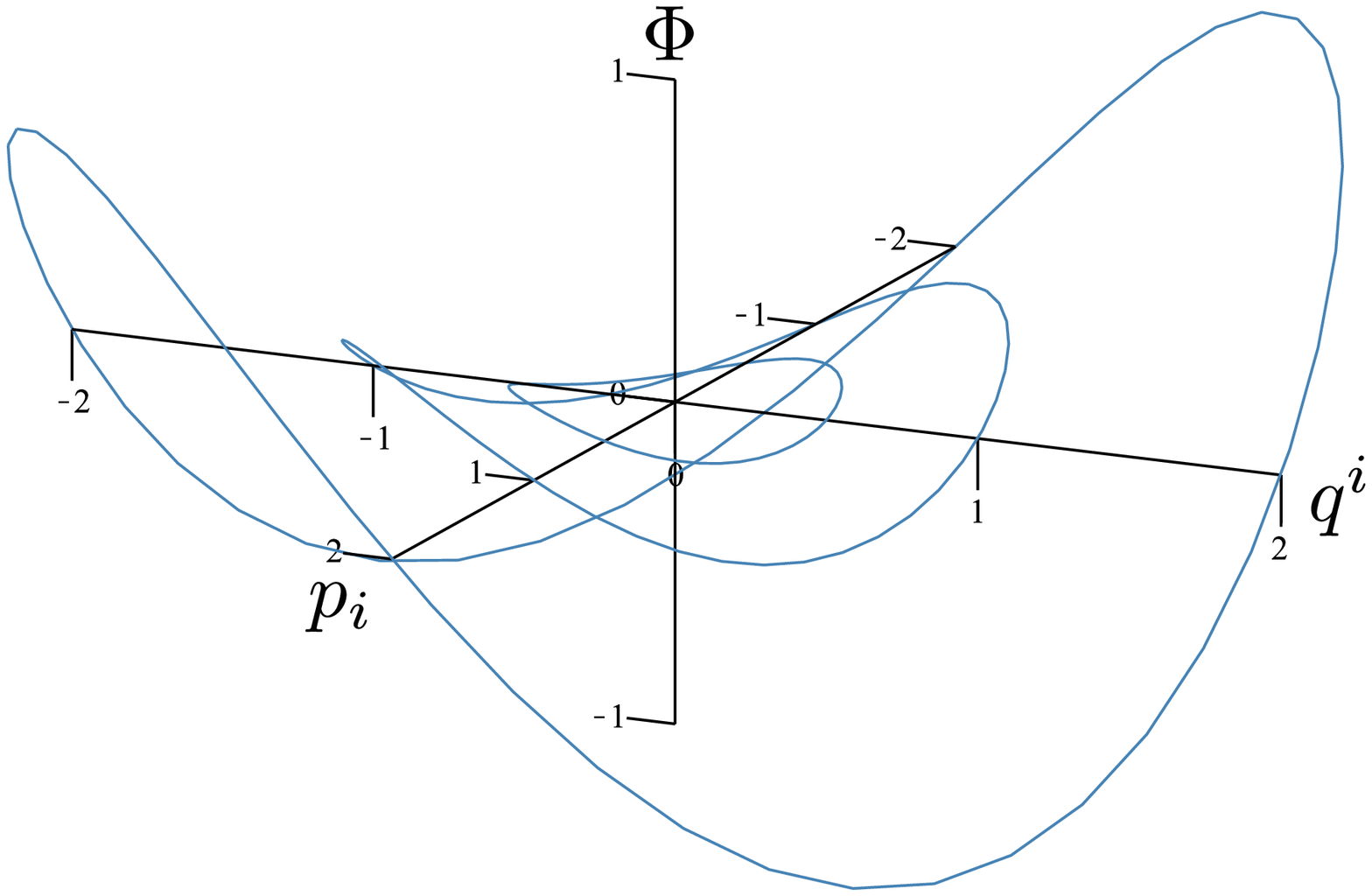}
\caption{Orbits of the infinitesimal Legendre symmetry generator $X_{L_i}$. On the left, we show the projected orbits on the $(q^i,p_i)$-plane for the initial conditions $(2,0,0)$, $(1,0,0)$ and $(1/2,0,0)$. On the right, we show the shape of the orbits of $X_{L_i}$ on the space $(q^i,p_i,\Phi)$ for the same initial conditions}
\label{fig1}
\end{figure}

Consider the contact Hamiltonian given by
	\beq
	\label{hamiltonian}
	h =  \sum_{i=1}^n\frac{1}{2} \left({q^i}^2 + {p_i}^2 \right),
	\eeq
then, its associated vector field is
	\beq
	\label{XL}
	X_{L} = \left[\sum_{i=1}^n\frac{1}{2} \left({q^{i}}^2 - {p_{i}}^2 \right) \right]\frac{\partial}{\partial \Phi} + q^{a} \frac{\partial }{\partial p_{a}} - p_{a} \frac{\partial}{\partial q^{a}}.
	\eeq
It is straightforward to obtain the flow of $X_L$. To acquire an intuition of the orbits of the contact Hamiltonian \eqref{XL}, let us consider the simpler generating Hamiltonian
	\beq
	\label{hams}
	h_i = \frac{1}{2} \left({q^i}^2  + {p_i}^2\right). 
	\eeq
Note that the Hamiltonian \eqref{hamiltonian} is simply the sum over $i$ of the individual functions \eqref{hams} and, therefore, the Hamiltonian vector field \eqref{XL} will be simply the sum of each Hamiltonian vector field 
	\beq
	X_{L_i} = \frac{1}{2} \left({q^{i}}^2 - {p_{i}}^2 \right)\frac{\partial}{\partial \Phi} + q^{i} \frac{\partial }{\partial p_{i}} - p_{i} \frac{\partial}{\partial q^{i}} \quad \text{(no sum over $i$)}.
	\eeq
To find the contact transformation generated by $X_{L_i}$, we need to integrate the flow
	\beq
	\frac{\d }{\d t} Z^A_i = X_{L_i}^a,
	\eeq
where the $Z^A$ are the Darboux coordinates of $\mathcal{T}$. Writting up explicitly the equations for the coordinate transformation we have
	\begin{align}
	\label{z1}
	\frac{\d}{\d t} \Phi_{(i)} 	&= \frac{1}{2} \left[ {q^{i}}^2 - {p_{i}}^2\right],\\
	\label{z2}
	\frac{\d }{\d t} p_{i} &= q^{i},\\
	\label{z3}
	\frac{\d }{\d t} q^{i} &= -p_{i}.
	\end{align}
These are the contact equivalent to Hamilton's equations. The last two, equations \eqref{z2} and \eqref{z3}, can be integrated immediately to obtain
	\begin{align}
	p_{(i)}(t) 	& = p_{(i)} \cos(t) + q^{(i)} \sin(t),\\
	q^{(i)}(t) 	& = -p_{(i)} \sin(t) + q^{(i)} \cos(t),
	\end{align}
and substituting these into \eqref{z1}, we find
	\beq
	\Phi_{(i)}(t) = \frac{1}{2} \left[ {q^{i}}^2 - {p_{i}}^2\right] \sin(t)\cos(t) - p_{(i)} q^{(i)} \sin^2(t) + \Phi \quad \text{(no sum over $i$)}.
	\eeq
Note that 	the value of $\Phi_{(i)}(t)$ is not fixed along the orbit of $X_{L_i}$ (see figure \ref{fig1}). Given an initial condition on the $q^i$ axis where the initial value $\Phi(0) = \Phi$ is a constant, there are exactly four points along the orbit of $X_{L_i}$ with the same value of $\Phi$, each corresponding to a $n \pi/2$ rotation in the $\left(q^{i},p_{i}\right)$-plane. Thus, we see that a $\pi/2$ rotation yields the partial Legendre transformation interchanging  the $i$th pair of conjugate variables,
	\begin{align}
	\label{transp}
	\tilde \Phi_{(i)} 	\equiv & \Phi_{(i)}\left(\frac{\pi}{2} \right) = \Phi - p_i q^i \quad \text{(no sum over $i$)},\\
	\tilde p_i 			\equiv & p_i\left(\frac{\pi}{2} \right) = q^i,\\
	\tilde q^i			\equiv & q^i\left(\frac{\pi}{2} \right) = - p_i.
	\end{align}

In the rest of this work, we will consider the generator for the total Legendre transformations, which interchanges every pair of conjugate variables and the expression for the Legendre transformed potential $\tilde \Phi$, equation \eqref{transp} above, will have an implicit sum over all the indices $i$.
	
\section{Infinitesimal Legendre Isometries}

The aim of this work is to obtain a class of metrics which remain invariant along the flow of the Hamiltonian vector field $X_L$ generating total Legendre transformations. For simplicity, let us consider the particular case with two degrees of freedom. All the results in this section can be easily extended to the general case and shall be presented elsewhere. Thus, consider the metric for the five-dimensional thermodynamic phase-space 
	\beq
	\label{metric}
	G = G_{AB}(Z^C)\ \d Z^A \otimes \d Z^B,
	\eeq
where the capital indices are taken over the set $\{1 ... 5\}$. As before, we work in \emph{ordered} Darboux coordinates, therefore, $Z^A \in \{\Phi,q^a,p_a \}$ for each $A$. 

From the symmetry of the metric tensor, there are 15 independent components in \eqref{metric} that must be determined by solving the Killing equation 
	\begin{equation}
	\dlie_{X_L} G = 0\,.	
	\label{legendresymm}
	\end{equation}
This system of equations has a freedom along the direction of $\Phi$ which can be removed once a fundamental representation is chosen (c.f. reference \cite{conformal}). Thus, choosing a contact 1-form -- $\eta$ -- in the class generating the contact structure, together with its associated Reeb vector field $R_\eta$ [c.f equation \eqref{reeb}, above], we see that $\eta$ is dragged along the orbits of the Reeb flow. We will demand that the metric \eqref{metric} shares this symmetry. A metric satisfying the equation
	\beq
	\lie_{R_\eta} G  = 0,
	\eeq
is called a $K$-contact. This imposes the restriction on the metric components of being independent of the coordinate $\Phi$. Therefore, instead of solving the general Legendre-Killing system, equation \eqref{legendresymm}, we will find the family of  $K$-contact metrics with Legendre symmetries obtained from
	\beq
	\label{kill2}
	\lie_{X_L} \mathcal{G} = 0,
	\eeq
where the $K$-contact $\mathcal{G}$ has the form
	\beq
	\label{metric2}
	\mathcal{G} = \mathcal{G}_{AB}(q^a,p_a)\ \d Z^A \otimes \d Z^B.
	\eeq
The form of the metric $\mathcal{G}$ remains quite general. Following the standard formulation of contact metric manifolds \cite{blair}, a contact metric is typically written in blocks. Thus, let us assume that the structure of $\mathcal{G}$ is of the form
	\beq
	\label{metric2.5}
	\mathcal{G} = \eta \otimes \eta + 2 \Omega_a^{\ b}(q^c,p_c) \ \d q^a \otimes \d p_b,
	\eeq
where we have fixed six out of fifteen independent components of \eqref{metric2} and we only need to determine the nine remaining functions $\Omega_a^{\ b}(q^c,p_c)$. This form of the metric is very restrictive with respect to the system \eqref{kill2}, and yields a unique class of solutions of the form
	\beq
	\label{metric3}
	\mathcal{G} = \eta \otimes \eta + 2 \Omega(q^c,p_c)\ \epsilon_a^{\ b}\ \d q^a \otimes \d p_b,
	\eeq
where $\Omega(q^a,p_a)$ is a non-vanishing, manifestly Legendre invariant function, i.e. it is a non-trivial solution to the constraint
	\beq
	\label{constraint}
	\{h, \Omega\} \equiv \cp{a}\prtf{\Omega}{\cx{a}}-\cx{a}\prtf{\Omega}{\cp{a}} =0 \,,
	\eeq
and $\epsilon_a^{\ b}$ are the components of the symplectic matrix
	\beq
	\epsilon = \left[\begin{array}{cc}
						0 & 1 \\
						-1 & 0
					\end{array} \right].
	\eeq
This result is relevant since it excludes all the metric tensors used in the GTD programme. The similarity of \eqref{metric3} with the GTD metrics, equations \eqref{quevedoGII} and \eqref{quevedoGIII}, is deceiving since the arrays $\xi^a_{\ b}$ and $\chi^a_{\ b}$ are diagonal, whereas $\epsilon_a^{\ b}$ is not. Thus, one can question the invariance of the GTD metrics altogether. One can see here that discrete invariance does not imply infinitesimal invariance, whilst the converse is true. Indeed, the GTD metrics change along the orbits of $X_L$ but recover their initial value every $\pi/2$ rotation.

This structure yields results which are significantly different with those stemming from the standard metric approaches for geometric thermodynamics.

\subsection{Thermodynamic interpretation: Revisiting the ideal gas}

Let us consider the equilibrium space as defined by the relations \eqref{equil}. In this case, the induced metric -- $g = \varphi^*\mathcal{G}$ -- is
	\beq
	g = 2\ \Omega\ \epsilon_a^{\ b}\ \Phi_{,bc}  \ \d q^a \otimes \d q^c,
	\eeq
where $\Phi_{,ab}$ stands for the second derivatives of the thermodynamic potential $\Phi$ with respect to the coordinates of $\mathcal{E}$.

In the particular case of the molarised ideal gas in the entropy representation, let us choose  $q^1 = u$ the molar energy and $q^2 = v$ the molar volume.  The potential is written as
	\beq
	s(u,v) = c_{v} \ln\left(u \right) + \log(v). 
	\eeq
Thus, the metric takes the simple, non-diagonal form
	\beq
	g_{{\rm ig}}= 2\ \Omega \left(\frac{c_v}{u^2} - \frac{1}{v^2} \right)\ \d u \otimes \d v,
	\eeq
and its determinant is given by
	\beq
	\det\left(g_{\rm ig} \right) = -\Omega^2  \left[\frac{\left(c_v v^2 - u^2\right)^2}{\left(u v\right)^4} \right].
	\eeq
	
\begin{figure}
\includegraphics[width=0.45\columnwidth]{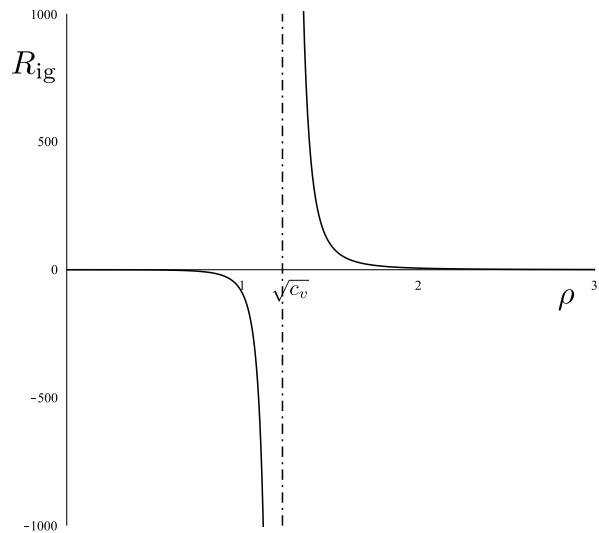}
\caption{Scalar curvature of the space of equilibrium states for the ideal gas. Note that if the metric function $\Omega$ is constant, the scalar curvature depends only on one parameter, the energy density $\rho$. Here we have used the heat capacity for a monatomic gas, $c_v=3/2$, and $\Omega=1$.}
\label{fig2}
\end{figure}

It is straightforward to compute the curvature scalar in terms of the equilibrium space coordinates $u$ and $v$. However, it is more convenient to write it up in terms of the energy density $\rho = u/v$ so that
	\beq
	\label{RS}
	R_{\rm ig} = \frac{2 \rho^2}{\Omega^3} \left[\frac{v^2 \left(\Omega \Omega_{,uv} - \Omega_{,u} \Omega_{,v} \right)}{\rho^2 - c_v}  + \frac{4 \Omega^2 c_v \rho}{\left(\rho^2 - c_v\right)^3} \right].
	\eeq
Note that if we demand that such a scalar vanishes, we have an extra equation for $\Omega$, apart from \eqref{constraint}. However, such a system is inconsistent. Moreover, assuming that the Legendre invariant function $\Omega$ (as a function of $\rho$) is regular for values of $\rho>0$, we observe that the curvature scalar \eqref{RS} is singular when the value of the energy density coincides with the square root of the heat capacity. Beyond that point, i.e. for values of $\rho>\sqrt{c_v}$, the curvature scalar decays very rapidly to zero. Thus, for this type of metrics, one can no longer keep the interpretation associating the curvature of the space of equilibrium states as a measure of thermodynamic interaction without any restriction, as it is normally done in every other metric programme for geometric thermodynamics. It is interesting that the curvature of the class of metrics we presented here exhibits a limit of applicability, presumably related with the thermodynamic limit itself (see figure \ref{fig2}).

\section{Conclusions}

In this work we have obtained a class of metrics for the thermodynamic phase-space $\mathcal{T}$ that are carried along the orbits of the infinitesimal generator of the total Legendre transformations in the case of two thermodynamic degrees of freedom. We have restricted the analysis to metrics which are fully compatible with the choice of one-form $\eta$ in the class generating the contact structure of $\mathcal{T}$. We do this by demanding that the metric has the Reeb vector field associated with $\eta$ as a symmetry. Such metrics are called $K$-contacts. The Legendre-Killing system resulting from \eqref{kill2} is a set of fifteen independent first order partial differential equations. By imposing the usual block decomposition for the metric of Riemannian contact manifolds [c.f. equation \eqref{metric2.5}], we fix six of the fifteen unknowns and the remaining nine are solved to yield a unique class of solutions of a single manifestly Legendre invariant function $\Omega(q^a,p_a)$. 

The solution obtained here, excludes all the metrics used in various programmes of geometric thermodynamics. This should be expected since, in general, discrete invariance does not imply infinitesimal symmetry. However, the class presented here stems from a differential system that probes the geometric structure of the thermodynamic phase-space consistently with the choice of contact 1-form. 

We revisited the space of equilibrium states for the ideal gas. We observed that we can no longer interpret the curvature of $\mathcal{E}$ as a measure of thermodynamic interaction since such requirement is inconsistent with the constraint that the metric function $\Omega$ should satisfy. Moreover, rewriting the scalar curvature in terms of the energy density of the system, it becomes singular when the numerical value of the energy density coincides with the square root of the heat capacity at constant volume. This might be an indication of the domain of applicability of the formalism in terms of the thermodynamic limit. This issue remains to be explored.

\section*{Acknowledgments}
DGP was funded by a CONACYT Scholarship. CSLM acknowledges financial support of a DGAPA-UNAM Post-doctoral Fellowship. The authors are thankful to Alessandro Bravetti and Francisco Nettel for helpful comments.


\end{document}